\documentclass[aps,prl,twocolumn,showpacs,superscriptaddress,longbibliography]{revtex4-2}
\usepackage{amsfonts}
\usepackage{amsmath}
\usepackage{graphicx}
\usepackage{bm}
\usepackage{amssymb}
\usepackage{dcolumn}
\usepackage{color}
\usepackage{soul}
\usepackage{multirow}
\usepackage{booktabs}
\usepackage{diagbox}
\usepackage{CJKutf8}
\usepackage{verbatim}
\usepackage[colorlinks,
    linkcolor=blue,
    anchorcolor=blue,
    citecolor=blue,
    urlcolor=blue,
]{hyperref}

\begin{document}

    \title{Two-dimensional room temperature ferromagnetic semiconductors}

    \author{Jia-Wen Li}
    \affiliation{Kavli Institute for Theoretical Sciences, University of Chinese Academy of Sciences, Beijng 100049, China}

    \author{Gang Su}
    \email{gsu@ucas.ac.cn}
    \affiliation{Kavli Institute for Theoretical Sciences, University of Chinese Academy of Sciences, Beijng 100049, China}
    \affiliation{Physical Science Laboratory, Huairou National Comprehensive Science Center, Beijing 101400, China}
    \affiliation{School of Physical Sciences, University of Chinese Academy of Sciences, Beijng 100049, China}
    \affiliation{Institute of Theoretical Physics, Chinese Academy of Sciences, Beijing 100190, China}

    \author{Bo Gu}
    \email{gubo@ucas.ac.cn}
    \affiliation{Kavli Institute for Theoretical Sciences, University of Chinese Academy of Sciences, Beijng 100049, China}
    \affiliation{Physical Science Laboratory, Huairou National Comprehensive Science Center, Beijing 101400, China}

    \begin{abstract}
        To realize ferromagnetic semiconductors with high Curie temperature $T\rm_C$ is still a challenge in spintronics.
        Recent experiments have obtained two-dimensional (2D) room temperature ferromagnetic metals, such as monolayers MnSe$_2$ and Cr$_3$Te$_6$.
        In this paper, by the density functional theory (DFT) calculations, we proposed a way to obtain 2D high $T\rm_C$ ferromagnetic semiconductors through element replacement in these ferromagnetic metals.
        High $T\rm_C$ ferromagnetic semiconductors are predicted in the monolayers (Mn, D)Se$_2$ and (Cr, D)$_3$Te$_6$, where element D is taken as vacancy, 3d, 4d and 5d transition metal elements.
        For the concentrations of D from 1/9 to 1/3, there are about 10 ferromagnetic semiconductors with $T\rm_C$ above 200 K, including (Cr$_{5/6}$, W$_{1/6}$)$_3$Te$_6$ and (Cr$_{4/6}$, Mo$_{2/6}$)$_3$Te$_6$ with $T\rm_C$ above 300 K.
        In addition, Mn(Se$_{6/8}$, Sb$_{2/8}$)$_2$ is also predicted to be a 2D ferromagnetic semiconductor with $T\rm_C$ above 300 K.
        Considering the fast developments on fabrication and manipulation of 2D materials, our theoretical results propose a way to explore the high temperature ferromagnetic semiconductors from experimentally obtained 2D high temperature ferromagnetic metals through element replacement approach.
    \end{abstract}
    \maketitle


    \textcolor{blue}{{\em Introduction.}}---Due to the interesting properties, there are many promising applications of magnetic semiconductors \cite{Dietl2010,Ohno2010,Sato2010,Jungwirth2006,Dietl2014,Ohno1998,Kalita2023,Fang2023,Telegin2022,Holub2005,Fiederling1999,Solin2000,Ohno2000,Mitra2001,Bebenin2004,Li2019,Song2018,Li2019,Gorbenko2007,Goel2023,Cinchetti2008}, such as spin injection maser \cite{Cinchetti2008,Goel2023}, circular polarized light emitting diodes \cite{Holub2005}, magnetic diode and p-n junction \cite{Fiederling1999,Solin2000,Ohno2000,Mitra2001,Bebenin2004}, magnetic tunnel junction \cite{Li2019,Song2018,Li2019}, and spin valve structures \cite{Gorbenko2007}, etc.
    These applications require magnetic semiconductors with high Curie temperature $T{\rm_C}$ above room temperature.
    However, the $T{\rm_C}$ of ferromagnetic semiconductors show low $T{\rm_C}$ far below room temperature, largely limit their applications.

    The intrinsic three-dimensional ferromagnetic semiconductors in nature show low $T{\rm_C}$ \cite{Baltzer1966}.
    In 2017, the successful synthesis of two-dimensional (2D) van der Waals ferromagnetic semiconductors CrI$_3$ \cite{Huang2017} and Cr$_2$Ge$_2$Te$_6$ \cite{Gong2017} in experiments has attracted extensive attentions to 2D ferromagnetic semiconductors.
    According to Mermin-Wagner theorem \cite{Mermin1966}, the magnetic anisotropy is essential to produce long-range magnetic order in 2D systems.
    Recently, more 2D ferromagnetic semiconductors have been obtained in experiments, such as Cr$_2$S$_3$ \cite{Chu2019}, CrCl$_3$ \cite{Cai2019}, CrBr$_3$ \cite{Zhang2019}, CrSiTe$_3$ \cite{Achinuq2021} and CrSBr \cite{Lee2021}, where the $T\rm_C$ are far below room temperature.
    In addition, many ferromagnetic semiconductors with $T{\rm_C}$ above room temperature have been predicted based on theoretical calculations \cite{You2023,Dong2019,You2019,You2020,Huang2019a,Jiang2018a,Huang2018a,Li2023b,Chen2020a,Song2021}, while their synthesis remains a challenge.

    On the other hand, the 2D van der Waals
    ferromagnetic metals with high $T\rm_C$ have been obtained in recent experiments.
    For example, $T\rm_C$ = 140 K in CrTe \cite{Wang2020}, 300 K in CrTe$_2$ \cite{Zhang2021,Xian2022,Wang2024}, 344 K in Cr$_3$Te$_6$ \cite{Chua2021}, 160 K in Cr$_3$Te$_4$ \cite{Li2022b}, 280 K in CrSe \cite{Zhang2019a}, 300 K in Fe$_3$GeTe$_2$ \cite{Deng2018,Fei2018}, 270 K in Fe$_4$GeTe$_2$ \cite{Seo2020}, 229 K in Fe$_5$GeTe$_2$ \cite{May2019}, 380 K in Fe$_3$GaTe$_2$ \cite{Zhang2022,Chen2024}, 300 K in MnSe$_2$ \cite{O’hara2018}, etc.

    Doping is a widely used method to control properties of 2D materials \cite{Li2017,Shen2022,Jiang2023,Zhang2020,Kar2019,Tiwari2021,Zhang2021c,Smiri2021,Bhat2022,Cai2015,Zhong2023,Xiao2023a,You2023}.
    In addition to the extensive studies of the diluted magnetic semiconductors by doping in 3D non-magnetic semiconductors \cite{Ohno1998,Dietl2014}, there are some recent efforts to obtain 2D ferromagnetic semiconductors by doping in 2D non-magnetic semiconductors \cite{Li2017,Shen2022,Jiang2023,Zhang2020,Cai2015,Zhong2023}.
    There are also some theoretical studies on the magnetic properties of the 2D doped materials \cite{Kar2019,Tiwari2021,Zhang2021c,Smiri2021,Bhat2022,Zhu2022}.
    In addition, doping can also induce metal-semiconductor transitions in 2D metals, as reported in recent experiments \cite{Hu2021a,Deng2022,Chen2024,Xiao2023a}, and theoretical studies \cite{Zhu2022,Hashmi2014,Liu2022}.

    In contrast to the conventional diluted magnetic semiconductors by doping magnetic impurities into non-magnetic semiconductors, is it possible to obtain magnetic semiconductors by doping impurities into magnetic metals?
    Inspired by the recent experimental and theoretical progress of 2D materials, we argue that it becomes possible in 2D systems.
    Because two important conditions are satisfied in 2D systems now.
    First, the 2D ferromagnetic metals with high $T\rm_C$.
    It is satisfied with the help of great progress of 2D ferromagnetic metals in recent experiments.
    Second, the doping-induced metal-semiconductor transitions.
    It is also possible in 2D metals as reported in the recent experimental and theoretical studies.

    \begin{figure*}[phbpt]
        \centering
        \includegraphics[width=2.02\columnwidth]{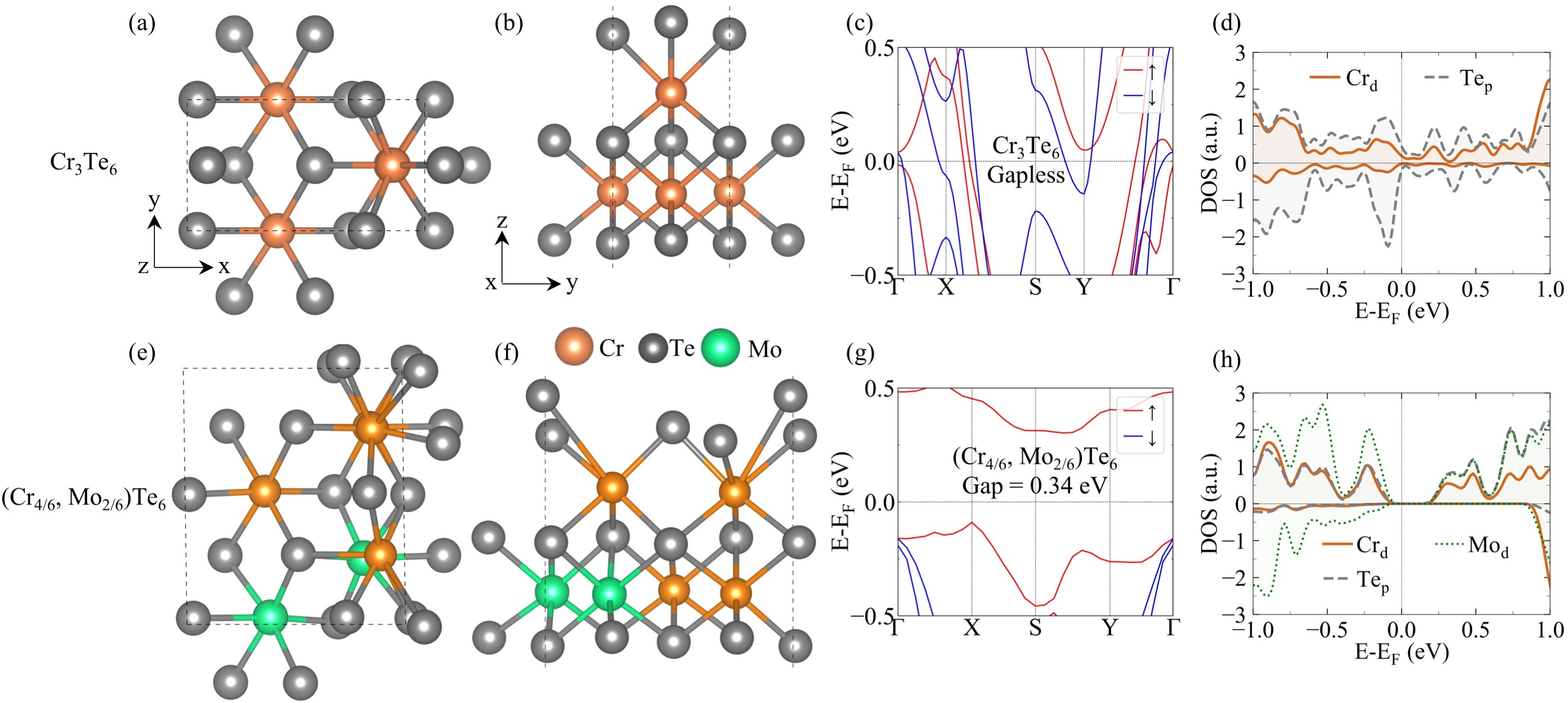}\\
        \caption{
            Crystal structures and band structures, partial density of state (PDOS) of monolayers Cr$_3$Te$_6$ and (Cr$_{4/6}$, Mo$_{2/6}$)$_3$Te$_6$.
            (a) Top view and (b) side view of monolayer Cr$_3$Te$_6$.
            (c) Spin polarized band structure of Cr$_3$Te$_6$.
            (d) Spin polarized partial density (PDOS) of monolayer Cr$_3$Te$_6$.
            (e) Top view and (f) side view of monolayer (Cr$_{4/6}$, Mo$_{2/6}$)$_3$Te$_6$.
            (g) Spin polarized band structure of (Cr$_{4/6}$, Mo$_{2/6}$)$_3$Te$_2$, showing a band gap of 0.34 eV.
            (h) Spin polarized PDOS of monolayer (Cr$_{4/6}$, Mo$_{2/6}$)$_3$Te$_6$.
            The band structures are obtained by the DFT calculation with HSE functional.
        }\label{Cr3Te6_band}
    \end{figure*}

    In this letter, we demonstrate the above idea to predict the ferromagnetic semiconductors with high $T\rm_C$ through element replacement in 2D ferromagnetic metals.
    For our purpose, the 2D ferromagnetic metals MnSe$_2$ with $T\rm_C$ = 300 K and Cr$_3$Te$_6$ with $T\rm_C$ = 344 K reported in recent experiments \cite{O’hara2018,Chua2021} are chosen as host metals.
    By density functional theory (DFT) calculations, the properties of monolayers (Mn$_{1-x}$, D$_x$)Se$_2$ and (Cr$_{1-x}$, D$_x$)$_3$Te$_6$ are studied, where D are vacancy, 3d, 4d and 5d transition metal elements, $x$ is concentration of D.
    For $x$ from 1/9 to 1/3, we have predicted about 10 ferromagnetic semiconductors with $T\rm_C$ above 200 K, i.e., (Mn$_{6/9}$, D$_{3/9}$)Se$_2$ with D as Mn vacancy, Au and Cd, (Cr$_{4/6}$, D$_{2/6}$)$_3$Te$_6$ with D as Ag, Hg and Mo.
    Their stability and feasibility of element replacement are confirmed by the calculations of formation energy, defect formation energy and molecular dynamics simulations.
    As an example, property of (Cr$_{4/6}$, Mo$_{2/6}$)$_3$Te$_6$ with $T\rm_C$ = 400 K and band gap of 0.34 eV is discussed in detail in the paper, and others are given in the Supplemental Material \cite{SM}.
    Our theoretical results propose a way to obtain high temperature ferromagnetic semiconductors by element replacement in 2D ferromagnetic metals in experiments.

    \textcolor{blue}{{\em Properties of monolayers Cr$_3$Te$_6$ and (Cr$_{4/6}$, Mo$_{2/6}$)$_3$Te$_6$.}}---The top and side views of crystal structure of monolayer Cr$_3$Te$_6$ are shown in Figs. \ref{Cr3Te6_band}(a) and \ref{Cr3Te6_band}(b), respectively, with a space group Pm (No. 6).
    The calculated in-plane lattice constants are $a_0=6.92$~\AA~and $b_0=3.82$~\AA, in agreement with experimental values of $a_0=6.9\pm0.1$~\AA~and $b_0=4.0\pm0.1$~\AA~\cite{Chua2021}.
    The spin polarized band structure and partial density of state (PDOS) of monolayer Cr$_3$Te$_6$ is obtained by the DFT calculation with HSE hybrid functional, as shown in Fig. \ref{Cr3Te6_band}(c) and \ref{Cr3Te6_band}(d), respectively.
    It is a ferromagnetic metal.

    To study the effect of element replacement in the monolayer (Cr, Mo)$_3$Te$_6$, a 1$\times$2$\times$1 supercell of Cr$_3$Te$_6$ is considered, including 6 Cr atoms.
    In the supercell, 1 and 2 of 6 Cr atoms are replaced by Mo atoms, corresponding to concentration $x$ of 1/6 and 2/6 in (Cr$_{1-x}$, Mo$_{x}$)$_3$Te$_6$, respectively.
    The top and side views of structure of monolayer (Cr$_{4/6}$, Mo$_{2/6}$)$_3$Te$_6$ is shown in Figs. \ref{Cr3Te6_band}(e) and \ref{Cr3Te6_band}(f), respectively, where 2 of 6 Cr atoms are replaced by Mo atoms.
    For the monolayer (Cr$_{1-x}$, Mo$_{x}$)$_3$Te$_6$ with a fixed $x$, all the possible configurations are considered, and the configuration with the lowest energy is taken as the most stable configuration.
    The detailed discussion on different configurations is given in Supplemental Material \cite{SM}.
    For the most stable configuration of monolayer (Cr$_{4/6}$, Mo$_{2/6}$)$_3$Te$_6$, as shown in Fig. \ref{Cr3Te6_band}(d), the optimized lattice constants are $a=6.85$~\AA~and $b=8.01$~\AA, corresponding to in-plane strain of -1.0\% and 4.7\% along x and y axis, respectively, for the host monolayer Cr$_3$Te$_6$.
    Significant structural distortion appears due to element replacement.
    The spin polarized band structure and PDOS of monolayer (Cr$_{4/6}$, Mo$_{2/6}$)$_3$Te$_6$ are obtained by the DFT calculation with HSE hybrid functional, as shown in Fig. \ref{Cr3Te6_band}(g) and \ref{Cr3Te6_band}(h).
    It is a ferromagnetic semiconductor with a band gap of 0.34 eV.
    Therefore, by replacing 2/6 of Cr atoms with Mo atoms in monolayer Cr$_3$Te$_6$, the ferromagnetic semiconductor (Cr$_{4/6}$, Mo$_{2/6}$)$_3$Te$_6$ is obtained.

    To study the structural stability, the formation energy $E\rm_{formation}$ of monolayer (Cr$_{1-x}$, Mo$_{x}$)$_3$Te$_6$ is calculated by
    \begin{align}
        E&{\rm_{formation}}= \nonumber\\
        &\frac{E{\rm_{(Cr_{1-x}, Mo_{x})_3Te_6}}
            -3(1-x)E{\rm_{\rm{Cr}}}-3xE{\rm_{\rm{Mo}}}-6E{\rm_{Te}}}{9},
        \label{eq:EF-2}
    \end{align}
    where $E{\rm_{(Cr_{1-x}, Mo_{x})_3Te_6}}$ is the energy of monolayer (Cr$_{1-x}$, Mo$_{x}$)$_3$Te$_6$.
    $E_{\rm{Cr}}$ and $E_{\rm{Mo}}$ are energies per atom for bulks of Cr and Mo with space group Im$\overline{3}$m.
    The formation energies are -0.84 eV/atom and -0.62 eV/atom for monolayers Cr$_3$Te$_6$ and (Cr$_{4/6}$, Mo$_{2/6}$)$_3$Te$_6$, respectively, indicating their stability.

    The defect formation energy $E\rm_{d.f.}$ of Mo in (Cr$_{1-x}$, Mo$_{x}$)$_3$Te$_6$ is given by \cite{Jiang2023}
    \begin{align}
        E{\rm_{d.f.}}
        =\frac{E{\rm_{(Cr_{1-x}, Mo_{x})_3Te_6}}-E{\rm_{Cr_3Te_6}}}{x}+3E{\rm_{\rm{Cr}}}-3E{\rm_{\rm{Mo}}},
        \label{eq:EF2}
    \end{align}
    where $E{\rm_{Cr_3Te_6}}$ is the energy of monolayer Cr$_3$Te$_6$, and the other terms are the same in Eq. \eqref{eq:EF-2}.
    The defect formation energy $E\rm_{d.f.}$ shows the change of energy of Cr$_3$Te$_6$ per Mo atom.
    The calculated defect formation energy of Cd in monolayer (Cr$_{1-x}$, Mo$_{x}$)$_3$Te$_6$ is -5.18 eV per Mo atom, which means that there is an energy reduction of 5.18 eV per Mo atom, and indicates the feasibility of replacing Cr atoms with Mo atoms in monolayer Cr$_3$Te$_6$.

    In addition, we performed molecular dynamics simulations of monolayer (Cr$_{1-x}$, Mo$_{x}$)$_3$Te$_6$ at 300 K, taking the NVT ensemble and running for 6 ps \cite{You2021a}.
    The results show that monolayer (Cr$_{1-x}$, Mo$_{x}$)$_3$Te$_6$ is thermodynamically stable \cite{SM}.

    \begin{figure}[phbpt]
        \centering
        \includegraphics[width=0.9\columnwidth]{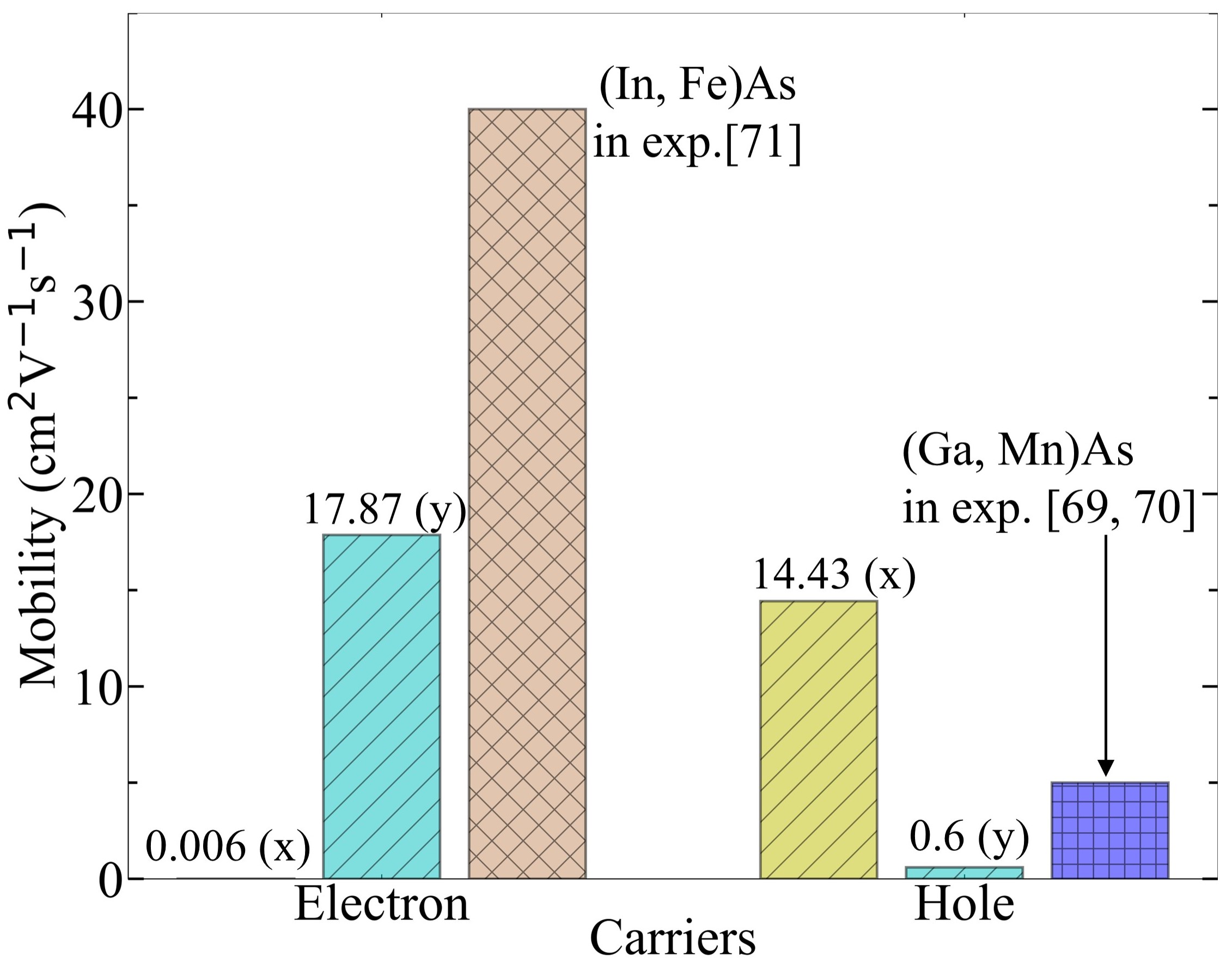}\\
        \caption{
            The calculated results of mobility for (Cr$_{4/6}$, Mo$_{2/6}$)$_3$Te$_6$ and the experimental results of hole mobility of (Ga, Mn)As \cite{Limmer2005,Moriya2003}, and electron mobility of (In, Fe)As \cite{Tanaka2014}.
        }\label{Mb}
    \end{figure}

    The mobility of monolayer (Cr$_{1-x}$, Mo$_{x}$)$_3$Te$_6$ were calculated based on deformation potential theory \cite{Schusteritsch2016}.
    The calculated results of mobility for (Cr$_{4/6}$, Mo$_{2/6}$)$_3$Te$_6$ by Perdew-Burke-Ernzerhof (PBE) exchange-correlation potential \cite{Perdew1996} is shown in Fig. \ref{Mb}, show an anisotropic mobility.
    Elections show a mobility of 17.87 cm$^2$V$^{-1}$s$^{-1}$ along y direction and a small mobility of 0.006 cm$^2$V$^{-1}$s$^{-1}$ along x direction.
    Holes show a mobility of 14.63 cm$^2$V$^{-1}$s$^{-1}$ along x direction and a small mobility of 0.6 cm$^2$V$^{-1}$s$^{-1}$ along y direction.
    Holes and electrons exhibit high mobility in x and y directions, respectively, while the mobility in another direction is nearly negligible.
    (Cr$_{4/6}$, Mo$_{2/6}$)$_3$Te$_6$ shows considerable mobility compared with FM semiconductors in experiments.
    The hole mobility of (Ga, Mn)As is about 10 cm$^2$V$^{-1}$s$^{-1}$ and the electron mobility in (In, Fe)As is tens of cm$^2$V$^{-1}$s$^{-1}$ \cite{Limmer2005,Moriya2003,Tanaka2014}.
    Details are shown in Supplemental Material \cite{SM}.


    \begin{figure*}[phbpt]
        \centering
        \includegraphics[width=2.1\columnwidth]{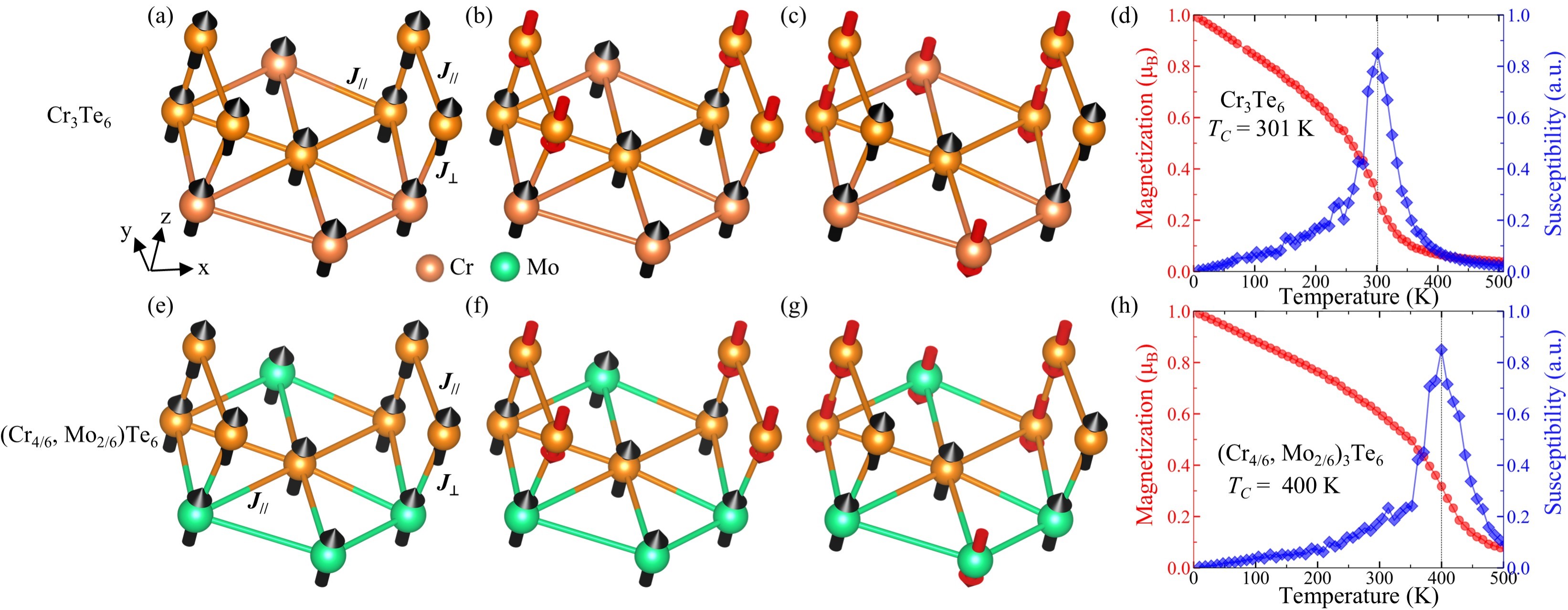}\\
        \caption{
            Spin configurations and magnetic properties of monolayers Cr$_3$Te$_6$ and (Cr$_{4/6}$, Mo$_{2/6}$)$_3$Te$_6$.
            A ferromagnetic (FM) (a), an antiferromagnetic (AFM) (b) and a ferrimagnetic (c) spin configurations are considered to get exchange coupling parameters J$_\parallel$ and J$_\perp$ of Cr$_3$Te$_6$.
            A ferromagnetic (FM) (e), an antiferromagnetic (AFM) (f) and a ferrimagnetic (g) spin configurations are considered for (Cr$_{4/6}$, Mo$_{2/6}$)$_3$Te$_6$.
            Magnetization and susceptibility of Cr$_3$Te$_6$ (d) and (Cr$_{4/6}$, Mo$_{2/6}$)$_3$Te$_6$ (e) as functions of temperature, obtained by the Monte Carlo simulations based on the 2D Heisenberg model.
            $T\rm_C$ of monolayers Cr$_3$Te$_6$ and (Cr$_{4/6}$, Mo$_{2/6}$)$_3$Te$_6$ are calculated as 301 K and 400 K, respectively.
            It is noted that $T\rm_C$ = 344 K for monolayer Cr$_3$Te$_6$ in experiment \cite{Chua2021}.
        }\label{Cr3Te6_mag}
    \end{figure*}

    To study the magnetic properties of monolayer (Cr$_{4/6}$, Mo$_{2/6}$)$_3$Te$_6$, we consider a 2D Heisenberg-type Hamiltonian, which can be written as
    \begin{equation}
        H=J_{\parallel}\sum\limits_{<i,j>}^{{\rm interlayer}}\vec{S_i} \cdot \vec{S_{j}}
        +J{_\perp}\sum\limits_{<i,j>}^{{\rm intralayer}}\vec{S_i} \cdot \vec{S_{j}}
        +A\sum\limits_{i} S_{iz}^2,\label{eq:b1}
    \end{equation}
    where $\vec{S_i}$ and $\vec{S_j}$ are spin operators of magnetic atoms at site $i$ and $j$, respectively.
    $J_\parallel$ and $J_\perp$ are the exchange coupling constants between the nearest-neighboring interlayer and intralayer magnetic atoms, respectively.
    The single-ion magnetic anisotropy parameter parameter $A$ is obtained by $AS^2=(E_{\perp}-E_{\parallel})/(N_{\rm{Cr}}+N_{\rm{Mo}})$, where N$\rm_{\rm{Cr}}$ and N$\rm_{\rm{Mo}}$ are numbers of Cr and Mo atoms in (Cr$_{4/6}$, Mo$_{2/6}$)$_3$Te$_6$, respectively.
    $E_{\perp}$ and $E_{\parallel}$ are energies of (Cr$_{4/6}$, Mo$_{2/6}$)$_3$Te$_6$ with out-of-plane and in-plane magnetization, respectively.
    By DFT results, $AS^2$ is calculated to be 0.95 meV/atom.

    For the monolayer (Cr$_{4/6}$, Mo$_{2/6}$)$_3$Te$_6$, a FM, an AFM and a ferrimagnetic (FIM) spin configurations are considered, as shown in Figs. \ref{Cr3Te6_mag}(e), \ref{Cr3Te6_mag}(f) and \ref{Cr3Te6_mag}(g), respectively.
    Their energies can be expressed as
    $E{\rm_{FM}}$=$J_{\parallel}$ $(4S_{\rm{Cr}}^2+8S_{\rm{Cr}}S_{\rm{Mo}}+2S_{\rm{Mo}}^2)$+$J_{\perp}(S_{\rm{Cr}}^2+S_{\rm{Cr}}S_{\rm{Mo}}$+$E_0$, $E{\rm_{AFM}}$=$J_{\parallel}(-4S_{\rm{Cr}}^2-2S_{\rm{Mo}}^2)$
    +$J_{\perp}\left(S_{\rm{Cr}}^2+ S_{\rm{Cr}} S_{\rm{Mo}} \right)$+$E_0$, $E_{FIM}$=$J_{\parallel}(2S_{\rm{Cr}}^2+8S_{\rm{Cr}}S_{\rm{Mo}}+2S_{\rm{Mo}}^2)$-$J_{\perp}$$(S_{\rm{Cr}}^2+S_{\rm{Cr}}S_{\rm{Mo}})$+$E_0$,
    where $S\rm_{\rm{Cr}}$ and $S\rm_{\rm{Mo}}$ are magnetic moments of Cr and Mo atoms, respectively.
    $E_0$ is the energy part independent of spin configurations, which is included in the total energy of DFT results for (Cr$_{4/6}$, Mo$_{2/6}$)$_3$Te$_6$.
    The DFT results show $S\rm_{\rm{Cr}}=3.645~\mu{\rm_B}$ and $S\rm_{\rm{Mo}}=2.264~\mu{\rm_B}$ in the supercell (Cr$_{4/6}$, Mo$_{2/6}$)$_3$Te$_6$.
    The exchange coupling parameters of (Cr$_{4/6}$, Mo$_{2/6}$)$_3$Te$_6$ can be calculated by $J_\parallel=(E{\rm_{FM}}-E{\rm_{AFM}})/\left(8S_{\rm{Cr}}^2+8S_{\rm{Cr}}S_{\rm{Mo}}+4S_{\rm{Mo}}^2\right)$,
    $J_\perp=(E{\rm_{FM}}-E_{FIM})/\left(2S_{\rm{Cr}}^2+2S_{\rm{Cr}}S_{\rm{Mo}}\right)$.
    The DFT results with HSE hybrid functional for relative total energy of monolayer (Cr$_{4/6}$, Mo$_{2/6}$)$_3$Te$_6$ in FM, AFM and FIM states are 0, 684 and 304 meV, respectively.
    The exchange coupling constants can be calculated as $J_\parallel=-3.55$ meV, $J_\perp=-7.05$ meV.
    By the Monte Carlo simulation based on the 2D Heisenberg model in Eq. \eqref{eq:b1}, the temperature dependent magnetization and susceptibility of monolayer (Cr$_{4/6}$, Mo$_{2/6}$)$_3$Te$_6$ are calculated, as shown in Fig. \ref{Cr3Te6_mag}(h).
    The Curie temperature is estimated as $T\rm_C$ = 400 K, higher than that of monolayer Cr$_3$Te$_6$.
    The calculation results show that the monolayer (Cr$_{4/6}$, Mo$_{2/6}$)$_3$Te$_6$ is a ferromagnetic semiconductor with band gap of 0.34 eV and $T\rm_C$ of 400 K.

    For monolayer Cr$_3$Te$_6$, spin configurations in Figs. \ref{Cr3Te6_mag}(a)-(c) are considered.
    Taking $S_{\rm{Mo}}=S_{\rm{Cr}}$, we obtain
    $E{\rm_{FM}}= \left(14J_{\parallel}+2J_{\perp}\right) S_{\rm{Cr}}^2
    +E_0$,
    $E{\rm_{AFM}}=
    \left(-6J_{\parallel}+2J_{\perp}\right) S_{\rm{Cr}}^2
    +E_0$,
    $E_{FIM} = \left(14J_{\parallel}-2J_{\perp}\right) S_{\rm{Cr}}^2
    +E_0$,
    and $J_\parallel =(E{\rm_{FM}}-E{\rm_{AFM}})/(20S_{\rm{Cr}}^2)$,
    $J_\perp     =(E{\rm_{FM}}-E_{FIM})/(4S_{\rm{Cr}}^2)$.
    The DFT results with HSE hybrid functional for relative total energy of monolayer Cr$_3$Te$_6$ in FM, AFM and FIM states are 0, 454 and 273 meV, respectively, giving $J_\parallel S_{\rm{Cr}}^2=-22.7$ meV and $J_\perp S_{\rm{Cr}}^2=-68.1$ meV.
    The magnetic moment $S\rm_{\rm{Cr}}=3.50~\mu{\rm_B}$ in monolayer Cr$_3$Te$_6$, therefore the exchange couping constants of monolayer Cr$_3$Te$_6$ can be calculated as $J_\parallel=-1.85$ meV and $J_\perp=-5.55$ meV.
    The single-ion magnetic anisotropy energy of monolayer Cr$_3$Te$_6$ is calculated to be $AS^2=0.91$ meV/Cr.
    The Monte Carlo simulation result gives $T\rm_C$ = 301 K of monolayer Cr$_3$Te$_6$, as shown in Fig. \ref{Cr3Te6_mag} (d).
    It is noted that for the monolayer Cr$_3$Te$_6$, the calculated $T\rm_C$ is lower than the experimental $T\rm_C$ = 344 K \cite{Chua2021}.





    \begin{table*}[thb]
        \setlength{\tabcolsep}{1.5mm}
        \caption{
            The band gap, formation energy $E\rm_{formation}$, defect formation energy $E{\rm_{d.f.}}$, coupling constants, MAE and Curie temperature $T\rm_C$ for monolayers (Mn$_{1-x}$, D$_{x}$)Se$_2$ and (Cr$_{1-x}$, D$_{x}$)$_3$Te$_6$.
            Element D is considered as vacancy, 3d, 4d and 5d transitional metal elements with different magnetic moments $M\rm_D$, and only the results with band gap is shown in the table.
            The results are obtained by density functional theory calculations and Monte Carlo simulations.
        }
        {
            \scalebox{1}

            {

                \begin{tabular}{c|c|c|c|c|c|c|c|c|c|c|c}
                    \hline\hline
                    \multicolumn{4}{c|}{Ferromagnetic Semiconductors} &
                    \multicolumn{8}{c}{Properties}
                    \\
                    \hline
                    \makebox[0.09 \textwidth][c]
                    {\multirow{2}{*}{Monolayers}} &
                    \makebox[0.04 \textwidth][c]
                    {\multirow{2}{*}{$x$}} &
                    \multicolumn{2}{c|}{Element D} &
                    \multicolumn{2}{c|}{Gap (eV)}&
                    \makebox[0.06 \textwidth][c]{
                        $E\rm_{d.f.}$}&
                    \makebox[0.06 \textwidth][c]{
                        $E\rm_{formation}$}&
                    \multicolumn{2}{c|}{$J$ (meV)} &
                    \makebox[0.06 \textwidth][c]
                    {MAE} &
                    \makebox[0.05 \textwidth][c]{
                        \multirow{2}{*}{$T\rm_C$ (K)}}
                    \\
                    \cline{3-6}\cline{9-10}
                    & &
                    \makebox[0.06 \textwidth][c]{type} &
                    \makebox[0.06 \textwidth][c]{$M\rm_D$ ($\mu{\rm_B}$)}&
                    \makebox[0.06 \textwidth][c]{PBE}&
                    \makebox[0.06 \textwidth][c]{HSE}&
                    (eV/D)&
                    (eV/atom)&
                    \makebox[0.055 \textwidth][c]{$J_\parallel$}&
                    \makebox[0.055 \textwidth][c]{$J_\perp $}
                    & (meV)&
                    \\
                    \hline
                    \multirow{14}{*}{(Cr$_{1-x}$, D$_x$)$_3$Te$_6$}
                    & \multirow{3}{*}{1/6}
                    & W
                    & 1.16 & 0 & 0.15 & -12.04 & -0.67 & -3.37 & -4.85 & -1.92 & 362 \\
                    & & Tc
                    & 1.79 & 0 & 0.02 & -13.11 & -0.73 & -1.50 & -1.70 & 0.71 & 222 \\
                    & & Pd
                    & 0 & 0 & 0.02 & -13.15 & -0.73 & -2.72 & -3.23 & 0.55 & 245 \\
                    \cline{2-12}
                    & \multirow{11}{*}{2/6}
                    & Ta
                    &
                    \multirow{8}{*}{0}
                    & 0.14 & 0.50 & -5.98 & -0.66 & -1.52 & -2.53 & 0.79 & 71 \\
                    & & Hf
                    & & 0.09 & 0.42 & -7.58 & -0.84 & -1.49 & -5.23 & 1.06 & 103 \\
                    & & Pd
                    & & 0 & 0.15 & -5.87 & -0.65 & -2.82 & -0.10 & -0.47 & 230 \\
                    & & Pt
                    & & 0.17 & 0.45 & -6.07 & -0.67 & -1.50 & -2.48 & 1.50 & 71 \\
                    & & Ag
                    & & 0.13 & 0.35 & -4.91 & -0.55 & -2.64 & / & 0.91 & 217 \\
                    & & Au
                    & & 0 & 0.03 & -5.01 & -0.56 & -1.96 & -0.59 & -0.48 & 159 \\
                    & & Zr
                    & & 0 & 0.12 & -7.71 & -0.86 & -2.57 & -0.56 & 1.29 & 198 \\
                    & & Hg
                    & & 0.09 & 0.33 & 6.38 & 0.71 & -2.89 & / & 0.34 & 220 \\
                    \cline{3-12}
                    & & Tc
                    & 2.75 & 0.10 & 0.35 & -5.45 & -0.61 & -1.40 & -2.20 & 0.32 & 182 \\
                    & & Mo
                    & 2.26 & 0.06 & 0.34 & -5.56 & -0.62 & -2.95 & -6.16 & 1.74 & 400 \\
                    & & Ni
                    & 0.83 & 0 & 0.07 & -5.50 & -0.61 & -2.64 & -6.89 & 0.67 & 175 \\
                    \hline
                    \multirow{8}{*}{(Mn$_{1-x}$, D$_x$)Se$_2$}
                    & 2/9
                    & Mn Vac.
                    & 0 & 0 & 0.07 & -4.48 & -0.67 & -1.61 & / & -1.28 & 145 \\
                    \cline{2-12}
                    & \multirow{7}{*}{3/9}
                    & Mn Vac.
                    & \multirow{6}{*}{0} & 0.28 & 0.64 & -2.93 & -0.37 & -2.93
                    & \multirow{7}{*}{/}
                    & -0.20& 252 \\
                    & & Nb
                    & & 0 & 0.19 & -6.03 & -0.67 & -2.24 & & 0.13 & 194 \\
                    & & Cd
                    & & 0.25 & 0.56 & -3.53 & -0.39 & -3.30 & & 1.13 & 276 \\
                    & & W
                    & & 0.02 & 0.19 & -4.37 & -0.49 & -1.09 & & -0.14 & 94 \\
                    & & Au
                    & & 0 & 0.13 & -2.85 & -0.32 & -2.03 & & -1.07 & 92 \\
                    & & Hg
                    & & 0.09 & 0.38 & -2.77 & -0.31 & -3.08 & & 1.06 & 226 \\
                    \cline{3-12}
                    & & Re
                    & 1.62 & 0.33 & 0.33 & -3.62 & -0.40 & -0.90 & & -1.11 & 59 \\
                    \hline\hline
                \end{tabular}
            }}

        \label{tab:res}
    \end{table*}

    \textcolor{blue}{{\em Properties of monolayers (Cr, D)$_3$Te$_6$ and (Mn, D)Se$_2$.}}---In the similar way, we have studied the monolayers (Cr, D)$_3$Te$_6$ and (Mn, D)Se$_2$, where element D is taken as vacancy, 3d, 4d, 5d transitional metal elements.
    We considered a supercell of 1$\times$2$\times$1 Cr$_3$Te$_6$ with 6 Cr atoms.
    1 and 2 of 6 Cr atoms are replaced by element D, corresponding to concentration $x$ as 1/6 and 2/6 in (Cr$_{1-x}$, D$_{x}$)$_3$Te$_6$, respectively.
    HSE hybrid functional approach is used to obtain the band gaps and coupling constants~\cite{Heyd2003}.
    As shown in Tab. \ref{tab:res}, some high $T\rm_C$ ferromagnetic semiconductors are predicted, such as (Cr$_{5/6}$, D$_{1/6}$)$_3$Te$_6$ with D as W, Tc and Pd, and (Cr$_{4/6}$, D$_{2/6}$)$_3$Te$_6$ with D as Pd, Ag, Hg and Mo with $T\rm_C$ above 200 K.

    We considered a supercell of 3$\times$3$\times$1 MnSe$_2$ with 9 Mn atoms.
    1, 2 and 3 of 9 Mn atoms are replaced by element D, corresponding to concentration $x$ of 1/9, 2/9 and 3/9 in (Mn$_{1-x}$, D$_{x}$)Se$_2$, respectively.
    The obtained ferromagnetic semiconductors of (Mn, D)Se$_2$ are shown in Table~\ref{tab:res}, where element D is nonmagnetic.
    $J_{\parallel}$ in Table \ref{tab:res} is the nearest in-plane coupling constants in (Mn, D)Se$_2$.
    Some high $T\rm_C$ ferromagnetic semiconductors are predicted, such as monolayers (Mn$_{6/9}$, D$_{3/9}$)Se$_2$ with D as Mn vacancy, Cd and Au with $T\rm_C$ higher than 200 K.
    In addition, the band gap of (Mn$_{1-x}$, D$_{x}$)Se$_2$ with D as Mn vacancy increase form 0.07 eV as $x$ = 2/9 to 0.64 eV as $x$ = 3/9.
    This behavior has been observed in 2D semimetal gray arsenene, whose band gap increases with the increase of vacancy concentrations \cite{Hu2021a}.

    As shown in Tab. \ref{tab:res}, the obtained 2D ferromagnetic semiconductors show negative formation energies and defect formation energies, indicating their structural stability and feasibility of element replacement.
    Details of different configurations are given in Supplemental Material \cite{SM}.

    The results in Tab. \ref{tab:res} predict about 10 ferromagnetic semiconductors with $T\rm_C$ above 200 K, including (Cr$_{5/6}$, W$_{1/6}$)$_3$Te$_6$ and (Cr$_{4/6}$, Mo$_{2/6}$)$_3$Te$_6$ with $T\rm_C$ above 300 K through element replacement in 2D ferromagnetic metals MnSe$_2$ and Cr$_3$Te$_6$.

    The properties of monolayers Mn(Se, D)$_2$ and Cr$_3$(Te, D)$_6$ are also discussed.
    All results of Cr$_3$(Te, D)$_6$ are metallic.
    Mn(Se$_{6/8}$, Sb$_{2/8}$)$_2$ is found to be a ferromagnetic semiconductor with $T_{\rm C}$ of 330 K and band gap of 0.26 eV.
    In addition, Mn(Se$_{6/8}$, D$_{2/8}$)$_2$ are found to be AFM semiconductors when D is Si, Ge and Sn, with Neel temperatures $T_{\rm N}$ of 153 K, 224 K and 208 K, respectively.
    Mn(Se$_{5/8}$, D$_{3/8}$)$_2$ with D as Si is AFM semiconductor with $T_{\rm N}$ of 86 K, and FM semiconductors with $T_{\rm C}$ of 181 K and 94 K when D is Ge and Sn.
    The details are given in Supplemental Material \cite{SM}.

    \textcolor{blue}{{\em Conclusion.}}---Inspired by the recent experimental progress on 2D metals, such as high Curie temperature $T{\rm_C}$ above room temperature and doping induced metal-semiconductor transition, we have theoretically studied the possible high $T{\rm_C}$ ferromagnetic semiconductors by element replacement in 2D ferromagnetic metals.
    For the 2D materials (Mn, D)Se$_2$ and (Cr, D)$_3$Te$_6$ with D being vacancy, 3d, 4d and 5d transitional metal elements and concentration of D from 1/9 to 1/3, we found about 10 ferromagnetic semiconductors with $T\rm_C$ above 200 K, including (Cr$_{5/6}$, W$_{1/6}$)$_3$Te$_6$ and (Cr$_{4/6}$, Mo$_{2/6}$)$_3$Te$_6$ with $T\rm_C$ above 300 K.
    In addition, Mn(Se$_{6/8}$, Sb$_{2/8}$)$_2$ is also found to be a 2D ferromagnetic semiconductor with Tc above 300 K.
    The calculation results of formation energy and defect formation energy indicate their structural stability and feasibility of the element replacement approach.
    Our theoretical results demonstrate a way to obtain high temperature ferromagnetic semiconductors from experimentally obtained 2D high $T\rm_C$ ferromagnetic metals through element replacement, and will simulate further research on high temperature ferromagnetic semiconductors.

    \textcolor{blue}{{\em Acknowledgements.}}---This work is supported by National Key R\&D Program of China (Grant No. 2022YFA1405100), National Natural Science Foundation of China (Grant No. 12074378), Chinese Academy of Sciences (Grants No. YSBR-030, No. JZHKYPT-2021-08, No. XDB33000000).

    %

\end{document}